\documentclass[twocolumn,aps,prc,superscriptaddress,showpacs,floatfix]{revtex4}

\usepackage{array}
\usepackage{epsfig}
\usepackage{amsmath}
\usepackage{multirow}
\usepackage{graphicx}
\usepackage{amsmath}
\usepackage{feynmf}
\usepackage[normalem]{ulem}  
\usepackage[dvips]{color} 

\renewcommand\sout{\bgroup \color{red} \ULdepth=-.5ex \ULset}

\newcommand{\Ex}[2]{\ifmmode{#1\times10^{#2}}\else{$#1\times10^{#2}$}\fi}

\begin{document}
\title{Renormalization of dimension 6 gluon operators}

\author{HyungJoo Kim}\affiliation{Department of Physics and Institute of Physics and Applied Physics, Yonsei
University, Seoul 120-749, Korea}
\author{Su Houng~Lee}\affiliation{Department of Physics and Institute of Physics and Applied Physics, Yonsei
University, Seoul 120-749, Korea}
\date{\today}
\begin{abstract}
We identify the independent dimension 6 twist 4 gluon operators and calculate their renormalization in the pure gauge theory.   By constructing the renormalization group invariant combinations, we find the scale invariant condensates that can be estimated in nonperturbative calculations and used in QCD sum rules for heavy quark systems in medium.
\end{abstract}


\maketitle

\section{Introduction}

Understanding the changes of the matrix elements of the gluon operators near the critical temperature in QCD offers a useful picture on the nature of the QCD phase transition\cite{Boyd:1996bx}.  These can also be used in QCD sum rule analysis to understand the changes and melting of heavy quark system at finite temperature\cite{Furnstahl:1989ji,Morita:2007pt,Gubler:2011ua}.

In the pure gauge theory, the lowest dimensional operators are the scalar gluon condensate and the twist 2 gluon operator.
These dimension 4 operators can be reexpressed in terms of the electric condensate and the magnetic condensate.  The temperature dependence of these operators can be calculated directly from lattice calculation of the space time and space space elementary plaquette\cite{Lee:1989qj,Boyd:1996bx} or from combining the calculation of the energy density and pressure\cite{Lee:2008xp}.
The calculations show that while there is rapid change of the electric condensate across the phase transition temperature, the magnetic condensate changes very little\cite{Lee:2008xp,Manousakis:1986jh}.

Using the temperature dependence of the dimension 4 condensates as the input in the QCD sum rule approach for the heavy quark system,  $J/\psi$ and $\eta_c$ was found to undergo a rapid property change across the phase transition\cite{Morita:2007pt,Lee:2008xp} and to their dissociation\cite{Gubler:2011ua} slightly above  the critical temperature.   Moreover, it was also found that the free energy extracted from lattice calculation is the relevant potential to describe $J/\psi$ in a potential model\cite{Lee:2013dca}.   The extension to finite density also has interesting application\cite{Klingl:1998sr}.

To further understand the phase transition in terms of local operators and to expand the findings for the charmonium system by using QCD sum rule to dimension 6 level, we will identify the dimension 6 and twist 4 gluon operators and calculate their renormalization in the pure gauge theory.   The renormalization of scalar dimension 4 operators and scalar dimension 6 operators are well known\cite{Tarrach,Narison:1983kn}.  Our result completes the calculation of renormalization of all the dimension 6 gluon operator, hence will be  a first step toward identifying their mixing and thus a systematic analysis in the operator product expansion (OPE) of heavy quark correlation functions up to dimension 6\cite{SuHoungLee}.

In section II, we will identify the independent operators at dimension 6. In section III, we will renormalize these independent operators up to one loop order.  The scale invariant vacuum condensate will then be given in section IV.  Section V is a summary.
\section{Independent operators}
The gauge invariant dimension 6 operators are obtained by combining the covariant derivative $D_\mu$ and the field strength tensor $G_{\mu\nu}$.  To find the independent even parity operators, we use the Bianchi identity and symmetry property of the indices. Here, we start from the operators that are of the type $(D_{a}G_{bc})(D_{d}G_{ef})$; that is, multiplication of two covariant component each composed of a covariant derivative acting on the field strength tensor.

For the scalar operator, the indices `$abcdef$' have to become `$aabbcc$' type.  Considering the indices,  the covariant component $(DG)$ can be a term in one of the two types of the Bianchi identities. In the first case, the three indices `$abc$' are independent while in the second case, two indices are identical and summed over `$aab$'.
\begin{eqnarray}
&Type1 : D_{a}G_{bc}+D_{b}G_{ca}+D_{c}G_{ab}=0   , \nonumber
\\&Type2 : D_{a}G_{ab}+D_{b}G_{aa}+D_{a}G_{ba}=0. \label{bianchi1}
\end{eqnarray}
The scalar dimension 6 operator can be obtained from one of the terms in the product of the same type in Eq.~(\ref{bianchi1}).
Initially, four operators can be constructed from  $(Type1) \times (Type1)$, and one operator from $(Type2) \times (Type2)$.
 However, among the four types of operators coming from $(Type1) \times (Type1)$, using the symmetry property of the indices, one can show that only one operator is independent, irrespective of the order of their indices. Therefore, there exists two independent scalar operators. That is,
\begin{flalign}
D_{\alpha}G^a_{\mu \nu}D_{\alpha}G^a_{\mu \nu},\, D_{\mu}G^a_{\mu \alpha}D_{\nu}G^a_{\nu \alpha}.
\end{flalign}

Using the equation of motion, the second operator can be written in terms of quark operator, which vanishes in the pure gauge theory. Using higher dimensional Bianchi identity of the form  $[D,[D,G]]=0$, one can show that the usually quoted scalar operator can be obtained by combining the two independent operators.
\begin{small}
\begin{eqnarray}
g f^{abc} G^a_{\mu \nu} G^b_{\mu \alpha} G^c_{\nu \alpha} =
D_\mu G^{a}_{\mu \alpha} D_\nu G^{a}_{\nu \alpha} - \frac{1}{2} D_\alpha G^{a}_{\mu \nu} D_\alpha G^{a}_{\mu \nu}.
\end{eqnarray}
\end{small}

Similarly, for the spin 2 operators, the indices `$abcdef$' become `$abccdd$'; that is, $ab$ indices remain free while $cd$ indices are summed over. Then, there are four types of Bianchi identities that become relevant.
\begin{flalign}
&Type1: D_{a}G_{bc}+D_{b}G_{ca}+D_{c}G_{ab}=0  \nonumber
\\&Type2: D_{a}G_{cd}+D_{c}G_{da}+D_{d}G_{ac}=0 \nonumber
\\&Type3: D_{c}G_{dd}+D_{d}G_{dc}+D_{d}G_{cd}=0 \nonumber
\\&Type4: D_{a}G_{cc}+D_{c}G_{ca}+D_{c}G_{ac}=0.
\end{flalign}
The full operator can be obtained from one of the terms in the multiplication of  $Type1$ to $Type3$ and $Type2,Type4$ to themselves. In this case, one operator is obtained from $(Type1) \times (Type3)$, two operators from $(Type2) \times (Type2)$, and one from $(Type4) \times (Type4)$.
However, using the higher dimensional Bianchi identity, one can obtain a relation among the four operators. Hence, there are only three independent dimension 6 spin2 gluon operators. In this work, we will use the  following  set\cite{SuHoungLee}:
\begin{flalign*}
&scalar\, :\, f^{abc}G^a_{\mu\nu}G^b_{\mu \alpha}G^c_{\nu \alpha},\, D_{\mu}G^a_{\mu\alpha}D_{\nu}G^a_{\nu\alpha}
\\&spin2\, :\, D_{\beta}G^a_{\mu\nu}D_{\alpha}G^a_{\mu\nu},\,D_{\mu}G^a_{\alpha\mu}D_{\nu}G^a_{\beta\nu},\,D_{\beta}G^a_{\alpha\mu}D_{\nu} G^a_{\mu\nu}.
\end{flalign*}

On the other hand, using the equation of motion, we find that only two gluon operators of dimension 6 remain in the pure gauge theory.  These are  $f^{abc}G^a_{\mu\nu}G^b_{\mu \alpha}G^c_{\nu \alpha}$ and $D_{\beta}G^a_{\mu\nu}D_{\alpha}G^a_{\mu\nu}$.  The latter operator is proportional to $f^{abc}G^{a}_{\alpha \mu}G^{b}_{\beta \nu}G^{c}_{\mu \nu}$ with two spin indices $(\alpha \beta)$.  Introducing the color E and B fields, we find the off diagonal components are of the forms $E^a_{\parallel}B^b_{\perp}B^c_{\parallel}$ or $E^a_{\parallel}E^b_{\perp}B^c_{\parallel}$, the matrix elements of which vanish in the medium at rest due to rotational invariance.  Therefore, the two independent  dimension 6 operators in the pure gauge theory that remain and  that constitute  the diagonal components and the scalar operators   are $f^{abc}B^a\cdot(B^b \times B^c)$ and $f^{abc}B^a\cdot(E^b \times E^c)$.

\section{Renormalization}

The renormalization of scalar operators are reported in Ref.~\cite{Narison:1983kn}.  Here, we  will focus on the spin2 traceless(Twist4) part. We will use the three independent set as mentioned in the previous section after making the operators symmetric and traceless with respect to the two spin indices. We will therefore discuss the renormalization of the following three operators:
\begin{small}
\begin{flalign}
&O_1=D_{\beta}G^a_{\mu\nu}D_{\alpha}G^a_{\mu\nu}|_{ST},
\\&O_{2}=D_{\mu}G^a_{\alpha\mu}D_{\nu} G^a_{\beta\nu}|_{ST},
\\&O_{3}=D_{\beta}G^a_{\alpha\mu}D_{\nu} G^a_{\mu\nu}|_{ST},
\end{flalign}
\end{small}
where we have suppressed the external indices $\alpha ,\beta$  in the left hand side and $O_{\alpha\beta}|_{ST}$ means that $1/2(O_{\alpha\beta}+O_{\beta\alpha})-1/4g_{\alpha\beta}{\rm Tr}(O_{\alpha\beta})$.
First, we will study the renormalization of $O_1$ up to one loop order using the background field method with zero momentum insertion\cite{Abbott}.

To study the renormalization of the operator, we consider the following Green's functions with external fields,
\begin{small}
\begin{align}
\begin{split}
\langle A^a_\mu A^b_\nu A^c_\lambda O_1 \rangle =Z_{1,1}Z_A\langle A^a_\mu &A^b_\nu A^c_\lambda O_{1B} \rangle \\&+\sum_{j=2}^3 Z_{1,j}\langle A^a_\mu A^b_\nu A^c_\lambda O_{jB} \rangle.
\end{split}
\end{align}
\end{small}
Here, $A_\mu^a$ is the background gluon field and $Z_A$ the backgroundfield renormalization constant. $O_{1B}$ represents the bare operator with renormalized fields and coupling.

The diagrams that contribute to the renormalization are shown in FIG.1 with the Feynman rules given in FIG.2. The two gluon vertex comes from contraction with $A_{\mu}^a(p)A_{\nu}^b(q)$, the three gluon vertex comes from $A_{\mu}^a(p)A_{\nu}^b(q)A_{\lambda}^c(r)$, and the four gluon vertex comes from $A_{\mu}^a(p)A_{\nu}^b(q)A_{\lambda}^c(r)A_{\omega}^d(k)$.
\newline\newline
\begin{figure}[h]
  \centering
  \includegraphics[width=0.5\textwidth]{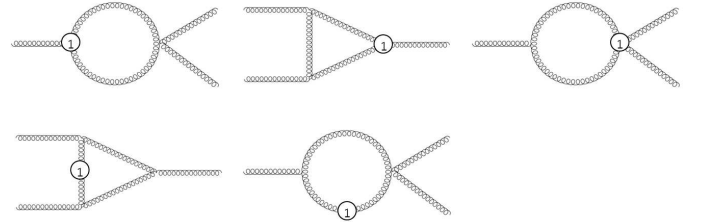}\\
  \caption{Diagrams contributing to the renormalization of $O_1$ to one loop order in the pure gauge theory.}
  \label{fig1}
\end{figure}
\begin{figure*}[h]
    \centering
    \begin{flushleft}
    \begin{minipage}[h]{.2\textwidth}
        \includegraphics[width=2.5cm]{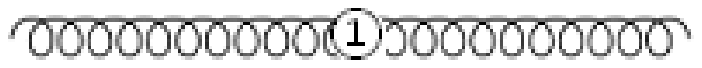}\hfill
        \label{fig2}
    \end{minipage}
        \begin{minipage}[h]{.7\textwidth}
        {\tiny
        \begin{flalign*}
        &4 p^{\alpha } p^{\beta } \delta _{ab} (p^2 g^{\mu \nu }-p^{\mu } p^{\nu })-\frac{1}{4} g^{\alpha \beta } (4 p^4 \delta _{ab} g^{\mu \nu }-4 p^2 p^{\mu } p^{\nu } \delta _{ab})
        \end{flalign*}
        }
    \end{minipage}
        \end{flushleft}
      \begin{flushleft}
    \begin{minipage}[h]{.2\textwidth}
        \includegraphics[width=2.5cm]{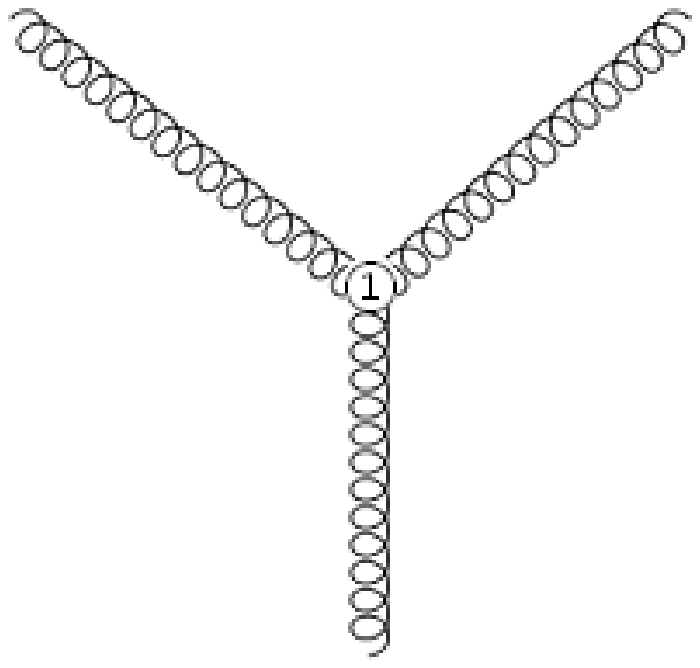}\hfill
        \label{fig2}
    \end{minipage}
        \begin{minipage}[h]{.7\textwidth}
        {\tiny
        \begin{flalign*}
        &-\frac{1}{2} i g f_{abc} (2 p^{\alpha } p^{\lambda } q^{\beta } g^{\mu \nu }-2 p^{\alpha } p^{\nu } q^{\beta } g^{\lambda \mu }+2 p^{\beta } q^{\alpha } (p^{\lambda } g^{\mu \nu }-p^{\nu } g^{\lambda \mu })-p^{\lambda } g^{\alpha \beta } g^{\mu \nu } p\cdot q
        +p^{\nu } g^{\alpha \beta } g^{\lambda \mu } p\cdot q+2 p^{\alpha } p^{\lambda } r^{\beta } g^{\mu \nu }+4 p^{\alpha } p^{\lambda } \\& r^{\mu } g^{\beta \nu }+4 p^{\beta } p^{\lambda } r^{\mu } g^{\alpha \nu }-2 p^{\alpha } p^{\nu } r^{\beta }
        g^{\lambda \mu }-2 p^{\lambda } p^{\nu } r^{\mu } g^{\alpha \beta }+2 p^{\beta } r^{\alpha }(p^{\lambda } g^{\mu \nu }-p^{\nu } g^{\lambda \mu })-p^{\lambda } g^{\alpha \beta } g^{\mu \nu } p\cdot r-4 p^{\alpha } g^{\beta \nu } g^{\lambda \mu } p\cdot r
        \\&-4 p^{\beta } g^{\alpha \nu } g^{\lambda \mu } p\cdot r+3 p^{\nu } g^{\alpha \beta } g^{\lambda \mu } p\cdot r )\text{ + (5 other terms from contraction order)}
        \end{flalign*}
        }
    \end{minipage}
    \end{flushleft}
    \begin{flushleft}
     \begin{minipage}[h]{.2\textwidth}
        \includegraphics[width=2.5cm]{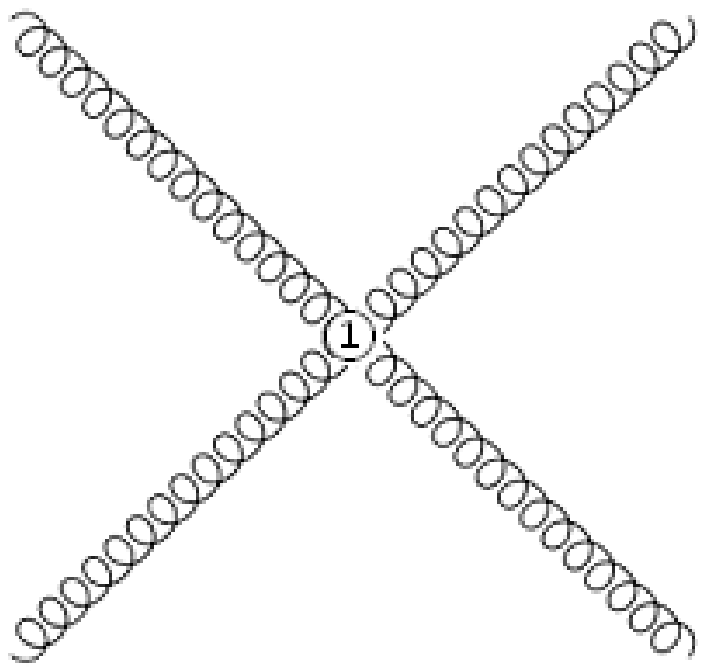}\hfill
        \label{fig2}
    \end{minipage}
           \begin{minipage}[h]{.7\textwidth}
        {\tiny
        \begin{flalign*}
        &-\frac{1}{4} g^2 f_{abx} f_{cdx} (4 k^{\alpha } p^{\beta } g^{\lambda \mu } g^{\nu \omega }+4 r^{\alpha } p^{\beta } g^{\lambda \mu } g^{\nu \omega }+4 k^{\alpha } q^{\beta } g^{\lambda \mu } g^{\nu \omega }+4 r^{\alpha } q^{\beta } g^{\lambda \mu } g^{\nu \omega }+4 k^{\alpha } g^{\beta \mu } q^{\lambda } g^{\nu \omega }+4 r^{\alpha } g^{\beta \mu } q^{\lambda } g^{\nu \omega }
        \\&+4 g^{\alpha \lambda } p^{\beta } k^{\mu } g^{\nu \omega }4 g^{\alpha \lambda } q^{\beta } k^{\mu } g^{\nu \omega }-g^{\alpha \beta }p^{\lambda } k^{\mu } g^{\nu \omega }-2 g^{\alpha \beta } q^{\lambda } k^{\mu } g^{\nu \omega }-g^{\alpha \beta } q^{\lambda } r^{\mu } g^{\nu \omega }-g^{\alpha \beta } g^{\lambda \mu } k\cdot p g^{\nu \omega }+8 g^{\alpha \lambda } g^{\beta \mu }
         \\&k\cdot q g^{\nu \omega }-3 g^{\alpha \beta } g^{\lambda \mu } k\cdot q g^{\nu \omega }-g^{\alpha \beta } g^{\lambda \mu } p\cdot r g^{\nu \omega }-g^{\alpha \beta } g^{\lambda \mu } q\cdot r g^{\nu \omega }-4 g^{\alpha \lambda } p^{\beta } g^{\mu \omega }k^{\nu }-4 g^{\alpha \lambda } q^{\beta } g^{\mu \omega } k^{\nu }+g^{\alpha \beta } p^{\lambda } g^{\mu \omega }
         \\&k^{\nu }+g^{\alpha \beta } q^{\lambda } g^{\mu \omega } k^{\nu }-2 g^{\alpha \beta } p^{\lambda } g^{\mu \omega } p^{\nu }
        +2 g^{\alpha \beta } g^{\lambda \mu } p^{\nu } p^{\omega }+4 p^{\alpha } g^{\beta \nu } (p^{\lambda } g^{\mu \omega }-g^{\lambda \mu } p^{\omega })+4 g^{\alpha \nu } p^{\beta } (p^{\lambda } g^{\mu \omega }-g^{\lambda \mu } p^{\omega })
         \\&-4 k^{\alpha } g^{\beta \mu } g^{\lambda \nu } q^{\omega }-4 r^{\alpha } g^{\beta \mu } g^{\lambda \nu } q^{\omega }+g^{\alpha \beta } g^{\lambda \nu } k^{\mu } q^{\omega }+g^{\alpha \beta } g^{\lambda \nu } r^{\mu } q^{\omega }-8 g^{\alpha \lambda }
         g^{\beta \mu } k^{\nu } q^{\omega }+2 g^{\alpha \beta } g^{\lambda \mu } k^{\nu } q^{\omega })\text{ + (23 other terms)}
        \end{flalign*}
        }
    \end{minipage}
    \end{flushleft}
    \begin{flushleft}
     \begin{minipage}[h]{.2\textwidth}
        \includegraphics[width=2.5cm]{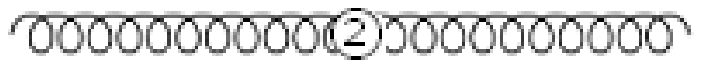}\hfill
        \label{fig2}
    \end{minipage}
           \begin{minipage}[h]{.7\textwidth}
        {\tiny
        \begin{flalign*}
        &\frac{1}{2} \delta _{ab} (p^2 (g^{\alpha \nu } (2 p^2 g^{\beta \mu }-2 p^{\beta } p^{\mu })+g^{\alpha \mu } (2 p^2 g^{\beta \nu }-2 p^{\beta } p^{\nu })+g^{\alpha \beta } (p^{\mu } p^{\nu }
        -p^2 g^{\mu \nu }))-2 p^{\alpha }(p^2(p^{\mu } g^{\beta \nu }+p^{\nu } g^{\beta \mu })-2 p^{\beta } p^{\mu } p^{\nu }))
        \end{flalign*}
        }
    \end{minipage}
     \begin{minipage}[h]{.2\textwidth}
        \includegraphics[width=2.5cm]{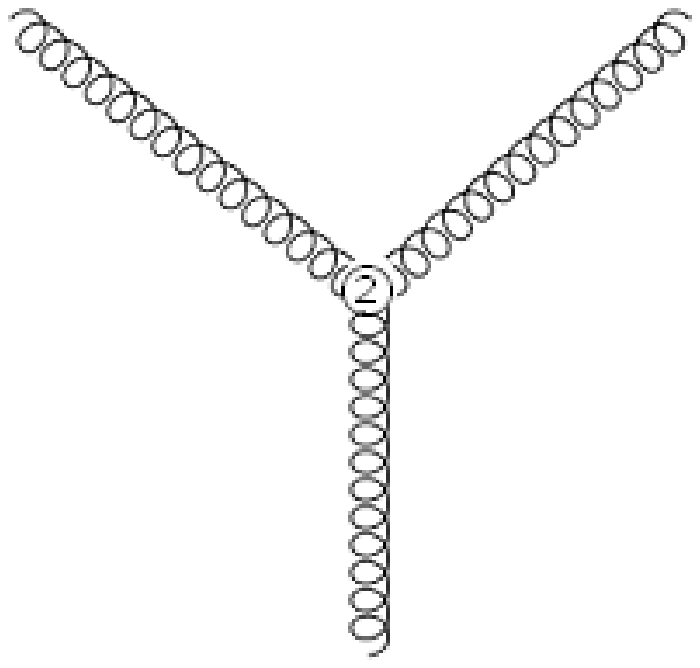}\hfill
        \label{fig2}
    \end{minipage}
           \begin{minipage}[h]{.7\textwidth}
        {\tiny
        \begin{flalign*}
        &\frac{1}{2} i g f_{abc} (2 p^{\beta } p^{\mu } q^{\lambda } g^{\alpha \nu }-p^{\mu } p^{\nu } q^{\lambda } g^{\alpha \beta }-2 p^2 q^{\lambda } g^{\alpha \nu } g^{\beta \mu }-2 p^2 q^{\lambda } g^{\alpha \mu } g^{\beta \nu }+p^2 q^{\lambda } g^{\alpha \beta } g^{\mu \nu }+2 p^{\alpha } p^{\mu }(g^{\beta \nu } (q^{\lambda }+r^{\lambda })+r^{\beta } g^{\lambda \nu }-r^{\nu }
        \\& g^{\beta \lambda })+2 p^{\beta } p^{\mu } r^{\lambda } g^{\alpha \nu }-p^{\mu } p^{\nu } r^{\lambda } g^{\alpha \beta }-2 p^{\beta } p^{\mu } r^{\nu } g^{\alpha \lambda }+p^{\lambda } p^{\mu } r^{\nu } g^{\alpha \beta }-2 p^2 r^{\beta } g^{\alpha \mu } g^{\lambda \nu }-2 p^2 r^{\lambda } g^{\alpha \nu } g^{\beta \mu }-2 p^2 r^{\lambda } g^{\alpha \mu } g^{\beta \nu }+p^2 r^{\lambda } \\&g^{\alpha \beta } g^{\mu \nu }+p^2 r^{\mu } g^{\alpha \beta } g^{\lambda \nu }+2 p^2 r^{\nu } g^{\alpha \mu } g^{\beta \lambda }+2 p^2 r^{\nu } g^{\alpha \lambda } g^{\beta \mu }-p^2 r^{\nu } g^{\alpha \beta } g^{\lambda \mu }+2 r^{\alpha } g^{\lambda \nu } (p^{\beta } p^{\mu }-p^2 g^{\beta \mu })-p^{\mu } g^{\alpha \beta } g^{\lambda \nu } p\cdot r )
        \\&\text{ + (5 other terms)}
        \end{flalign*}
        }
    \end{minipage}
    \end{flushleft}
    \begin{flushleft}
     \begin{minipage}[h]{.2\textwidth}
        \includegraphics[width=2.5cm]{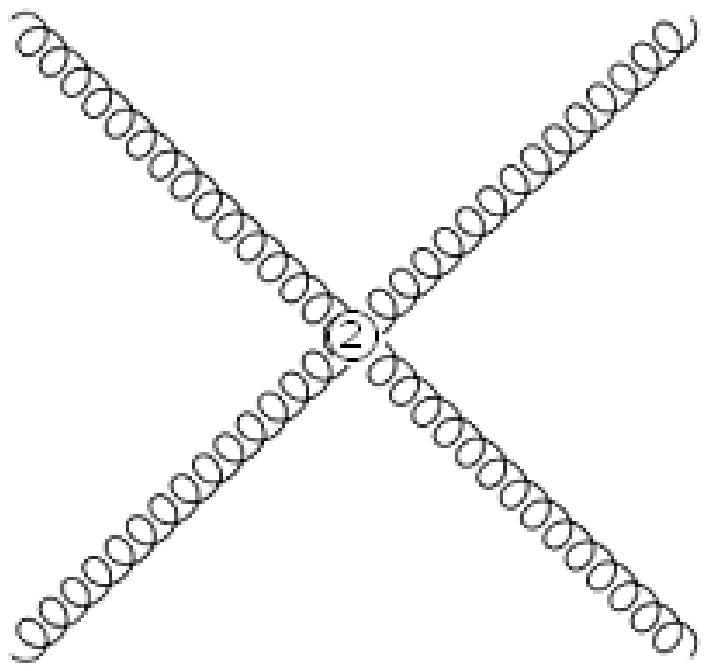}\hfill
        \label{fig2}
    \end{minipage}
           \begin{minipage}[h]{.7\textwidth}
        {\tiny
        \begin{flalign*}
        &-\frac{1}{4} g^2 f_{abx} f_{cdx} (4 p^{\alpha } g^{\beta \lambda } p^{\mu } g^{\nu \omega }+4 g^{\alpha \lambda } p^{\beta } p^{\mu } g^{\nu \omega }-2 g^{\alpha \beta } p^{\lambda } p^{\mu } g^{\nu \omega }-g^{\alpha \beta } k^{\lambda } q^{\mu }
         g^{\nu \omega }-4 g^{\alpha \mu } g^{\beta \lambda } p^2 g^{\nu \omega }-4 g^{\alpha \lambda } g^{\beta \mu } p^2 g^{\nu \omega }
         \\&+2 g^{\alpha \beta } g^{\lambda \mu } p^2 g^{\nu \omega }+g^{\alpha \beta } g^{\lambda \omega } q^{\mu } k^{\nu }+4 g^{\alpha \mu }k^{\beta } g^{\lambda \omega } p^{\nu }-4 g^{\alpha \mu } g^{\beta \omega } k^{\lambda } p^{\nu }+g^{\alpha \beta } k^{\lambda } g^{\mu \omega } p^{\nu }-g^{\alpha \beta } g^{\lambda \omega } k^{\mu } p^{\nu }+4 g^{\alpha \mu } k^{\beta } \\&g^{\lambda \omega }q^{\nu }-4 g^{\alpha \mu } g^{\beta \omega } k^{\lambda } q^{\nu }+g^{\alpha \beta } k^{\lambda } g^{\mu \omega } q^{\nu }-g^{\alpha \beta } g^{\lambda \omega } k^{\mu } q^{\nu }-g^{\alpha \beta } q^{\lambda } g^{\mu \nu } k^{\omega }+g^{\alpha \beta } g^{\lambda \nu } q^{\mu } k^{\omega }+4 g^{\alpha \mu } g^{\beta \lambda } p^{\nu } k^{\omega }-g^{\alpha \beta }
          \\&g^{\lambda \mu } p^{\nu } k^{\omega }+4 g^{\alpha \mu } g^{\beta \lambda } q^{\nu } k^{\omega }-g^{\alpha \beta } g^{\lambda \mu } q^{\nu }k^{\omega }+g^{\alpha \beta } k^{\lambda } g^{\mu \nu } q^{\omega }-g^{\alpha \beta } q^{\lambda } g^{\mu \nu } r^{\omega }+g^{\alpha \beta } g^{\lambda \nu } q^{\mu } r^{\omega }+4 g^{\alpha \mu } g^{\beta \lambda } p^{\nu } r^{\omega }
         \\&-g^{\alpha \beta } g^{\lambda \mu } p^{\nu } r^{\omega }+4 g^{\alpha \mu } g^{\beta \lambda } q^{\nu } r^{\omega }-g^{\alpha \beta } g^{\lambda \mu } q^{\nu } r^{\omega }+4 q^{\alpha } g^{\mu \nu } (k^{\beta } g^{\lambda \omega }-g^{\beta \omega }k^{\lambda }+g^{\beta \lambda } (k^{\omega }+r^{\omega }))-4 g^{\alpha \nu } q^{\mu } (k^{\beta } g^{\lambda \omega }
         \\&-g^{\beta \omega } k^{\lambda }+g^{\beta \lambda } (k^{\omega }+r^{\omega }))-g^{\alpha \beta } g^{\lambda \omega } g^{\mu \nu } k\cdot q )\text{ + (23 other terms)}
        \end{flalign*}
        }
    \end{minipage}
    \end{flushleft}
    \begin{flushleft}
     \begin{minipage}[h]{.2\textwidth}
        \includegraphics[width=2.5cm]{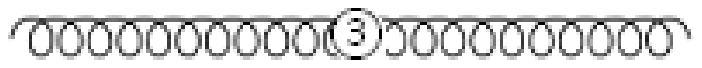}\hfill
        \label{fig2}
    \end{minipage}
           \begin{minipage}[h]{.7\textwidth}
        {\tiny
        \begin{flalign*}
        &\frac{1}{2} \delta _{ab} (4 p^{\alpha } p^{\beta }-p^2 g^{\alpha \beta }) (p^{\mu } p^{\nu }-p^2 g^{\mu \nu })
        \end{flalign*}
        }
    \end{minipage}
     \begin{minipage}[h]{.2\textwidth}
        \includegraphics[width=2.5cm]{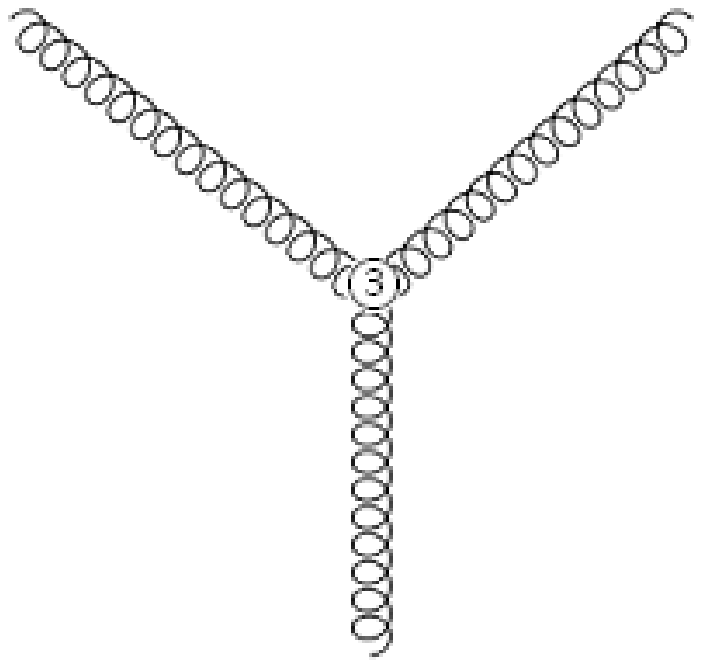}\hfill
        \label{fig2}
    \end{minipage}
           \begin{minipage}[h]{.7\textwidth}
        {\tiny
        \begin{flalign*}
        &\frac{1}{4} i g f_{abc} (4 r^{\alpha } g^{\beta \nu } p^{\lambda } p^{\mu }+2 g^{\alpha \nu } q^{\beta } p^{\lambda } p^{\mu }+4 g^{\alpha \nu } r^{\beta } p^{\lambda } p^{\mu }+g^{\alpha \beta } q^{\lambda } p^{\nu } p^{\mu }+g^{\alpha \beta } r^{\lambda } p^{\nu } p^{\mu }-g^{\alpha \beta } p^{\lambda } q^{\nu } p^{\mu }-3 g^{\alpha \beta } p^{\lambda } r^{\nu } p^{\mu }
        \\&-2 g^{\alpha \nu } g^{\beta \lambda } p\cdot r p^{\mu }-2 g^{\alpha \lambda } g^{\beta \nu } p\cdot r p^{\mu }+2 g^{\alpha \beta } g^{\lambda \nu } p\cdot r p^{\mu }-2 g^{\alpha \mu } p^{\beta } q^{\lambda } p^{\nu }-2 g^{\alpha \mu } p^{\beta } r^{\lambda } p^{\nu }+2 g^{\alpha \mu } p^{\beta } p^{\lambda } r^{\nu }-4 r^{\alpha } g^{\beta \nu } g^{\lambda \mu } p^2-2 g^{\alpha \nu }
        \\& q^{\beta } g^{\lambda \mu } p^2-4 g^{\alpha \nu } r^{\beta } g^{\lambda \mu } p^2-g^{\alpha \beta } q^{\lambda } g^{\mu \nu } p^2-g^{\alpha \beta } r^{\lambda } g^{\mu \nu } p^2+2 g^{\alpha \nu } g^{\beta \lambda } r^{\mu } p^2+2 g^{\alpha \lambda } g^{\beta \nu } r^{\mu } p^2-2 g^{\alpha \beta } g^{\lambda \nu } r^{\mu } p^2+g^{\alpha \beta } g^{\lambda \mu } q^{\nu } p^2+3 \\&g^{\alpha \beta } g^{\lambda \mu } r^{\nu } p^2+2 q^{\alpha } g^{\beta \nu } (p^{\lambda } p^{\mu }-g^{\lambda \mu } p^2)-2 g^{\alpha \mu } p^{\beta } g^{\lambda \nu } p\cdot r+2 p^{\alpha } (2 p^{\beta } (q^{\lambda } g^{\mu \nu }+r^{\lambda } g^{\mu \nu }
        \\&+g^{\lambda \nu } r^{\mu }-g^{\lambda \mu } r^{\nu })-g^{\beta \mu } (q^{\lambda } p^{\nu }+r^{\lambda } p^{\nu }-p^{\lambda } r^{\nu }+g^{\lambda \nu } p\cdot r)))\text{ + (5 other terms)}
        \end{flalign*}
        }
    \end{minipage}
    \end{flushleft}
    \begin{flushleft}
     \begin{minipage}[h]{.2\textwidth}
        \includegraphics[width=2.5cm]{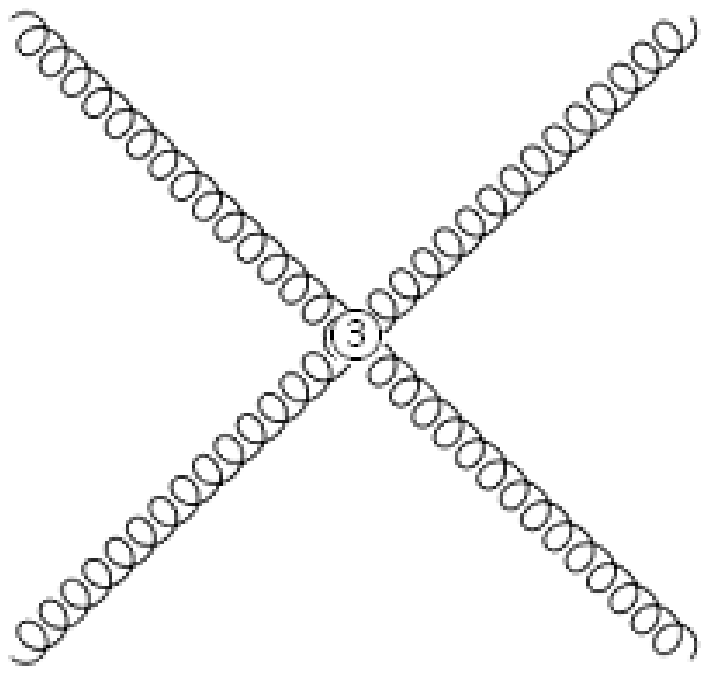}\hfill
        \label{fig2}
    \end{minipage}
           \begin{minipage}[h]{.7\textwidth}
        {\tiny
        \begin{flalign*}
        &\frac{1}{4} g^2 f_{abx} f_{cdx} (2 g^{\alpha \mu } p^{\beta } k^{\lambda } g^{\nu \omega }+4 g^{\alpha \mu } q^{\beta } k^{\lambda } g^{\nu \omega }+2 g^{\alpha \mu } p^{\beta } p^{\lambda } g^{\nu \omega }-g^{\alpha \beta } k^{\lambda } p^{\mu }g^{\nu \omega }
        -g^{\alpha \beta } p^{\lambda } p^{\mu } g^{\nu \omega }-2 g^{\alpha \beta } k^{\lambda } q^{\mu } g^{\nu \omega }+g^{\alpha \beta }
        \\&g^{\lambda \mu } p^2 g^{\nu \omega }-2 g^{\alpha \mu } p^{\beta } g^{\lambda \omega } k^{\nu }-4 g^{\alpha \mu } q^{\beta } g^{\lambda \omega } k^{\nu }+g^{\alpha \beta } g^{\lambda \omega } p^{\mu } k^{\nu }+2 g^{\alpha \beta } g^{\lambda \omega } q^{\mu } k^{\nu }-2 g^{\alpha \mu } p^{\beta } g^{\lambda \nu } k^{\omega }-4 g^{\alpha \mu } q^{\beta } g^{\lambda \nu } k^{\omega }
        \\&+2 g^{\alpha \nu } g^{\beta \mu } q^{\lambda } k^{\omega }+2 g^{\alpha \mu } g^{\beta \nu } q^{\lambda } k^{\omega }-g^{\alpha \beta } q^{\lambda } g^{\mu \nu } k^{\omega }+g^{\alpha \beta } g^{\lambda \nu } p^{\mu } k^{\omega }+2 g^{\alpha \beta } g^{\lambda \nu }q^{\mu } k^{\omega }-2 g^{\alpha \nu } g^{\beta \lambda } p^{\mu } p^{\omega }-2 g^{\alpha \lambda } g^{\beta \nu } \\&p^{\mu } p^{\omega }+g^{\alpha \beta } g^{\lambda \nu } p^{\mu } p^{\omega }-2 g^{\alpha \nu } g^{\beta \mu } k^{\lambda } q^{\omega }-2 g^{\alpha \mu } g^{\beta \nu } k^{\lambda } q^{\omega }+g^{\alpha \beta } k^{\lambda } g^{\mu \nu } q^{\omega }-2 g^{\alpha \mu } p^{\beta } g^{\lambda \nu } r^{\omega }-4 g^{\alpha \mu } q^{\beta } g^{\lambda \nu } r^{\omega }+2 g^{\alpha \nu } g^{\beta \mu }
        \\& q^{\lambda } r^{\omega }+2 g^{\alpha \mu } g^{\beta \nu } q^{\lambda } r^{\omega }-g^{\alpha \beta } q^{\lambda } g^{\mu \nu } r^{\omega }+g^{\alpha \beta } g^{\lambda \nu } p^{\mu } r^{\omega }+2 g^{\alpha \beta } g^{\lambda \nu } q^{\mu } r^{\omega }
        +4 q^{\alpha } g^{\beta \mu } (k^{\lambda } g^{\nu \omega }-g^{\lambda \omega } k^{\nu }-g^{\lambda \nu } (k^{\omega }+r^{\omega }))
        \\&-2 p^{\alpha } (2 p^{\beta }g^{\lambda \mu } g^{\nu \omega }+g^{\beta \mu } (-k^{\lambda } g^{\nu \omega }-p^{\lambda } g^{\nu \omega }+g^{\lambda \omega } k^{\nu }+g^{\lambda \nu } k^{\omega }+g^{\lambda \nu } r^{\omega }))+2 g^{\alpha \nu } g^{\beta \mu } g^{\lambda \omega } k\cdot q+2 g^{\alpha \mu } g^{\beta \nu } g^{\lambda \omega } k\cdot q-g^{\alpha \beta } g^{\lambda \omega } \\&g^{\mu \nu } k\cdot q+2 g^{\alpha \nu } g^{\beta \lambda } g^{\mu \omega } p^2+2 g^{\alpha \lambda } g^{\beta \nu } g^{\mu \omega } p^2-g^{\alpha \beta } g^{\lambda \nu } g^{\mu \omega } p^2)\text{ + (23 other terms)}
        \end{flalign*}
        }
    \end{minipage}
    \end{flushleft}
     \caption{Feynman rules in the background field method for the pure gauge theory.}
\end{figure*}

The calculation is performed using dimensional regularization $D=4-2\epsilon$ in Feynman gauge for SU(N).
The calculation for the  other operators involves the same diagrams and hence can be repeated similarly.  The following is  the collected result of the renormalization constants.
\begin{flalign}
&Z_{1,1}=1+\frac{3N}{4}\frac{\alpha_s}{\pi\epsilon}
\\&Z_{1,2}=-\frac{N}{12}\frac{\alpha_s}{\pi\epsilon}
\\&Z_{1,3}=\frac{2N}{3}\frac{\alpha_s}{\pi\epsilon}
\\&Z_{2,1}=0
\\&Z_{2,2}=1+\frac{N}{3}\frac{\alpha_s}{\pi\epsilon}
\\&Z_{2,3}=\frac{N}{24}\frac{\alpha_s}{\pi\epsilon}
\\&Z_{3,1}=0
\\&Z_{3,2}=\frac{N}{6}\frac{\alpha_s}{\pi\epsilon}
\\&Z_{3,3}=1+\frac{7N}{24}\frac{\alpha_s}{\pi\epsilon}.
\end{flalign}

\section{Scale invariant condensates}

The  scale invariant condenstates  can be obtained by diagonalization the following matrix $Z$.
\begin{align}
Z=\left(
\begin{array}{ccc}
 1+\frac{3 N \alpha_s }{4 \pi  \epsilon } & -\frac{N \alpha_s }{12 \pi  \epsilon } & \frac{2 N \alpha_s }{3 \pi  \epsilon } \\
 0 & 1+\frac{N \alpha_s }{3 \pi  \epsilon} & \frac{N \alpha_s }{24 \pi  \epsilon } \\
 0 & \frac{N \alpha_s }{6 \pi  \epsilon } & 1+\frac{7 N \alpha_s }{24 \pi  \epsilon }
\end{array}
\right).
\end{align}
We then  find the following new operator set, which corresponds to the eigenvectors of $Z$.
\begin{small}
\begin{flalign}
&\left\langle O_{1\text{new}}\right\rangle =\left\langle O_1\right\rangle
\\ &\left\langle O_{2\text{new}}\right\rangle =\left\langle \frac{-653+21\sqrt{17}}{424}O_1+\frac{1-\sqrt{17}}{8}O_2+O_3\right\rangle
\\ &\left\langle O_{3\text{new}}\right\rangle =\left\langle \frac{-653-21\sqrt{17}}{424}O_1+\frac{1+\sqrt{17}}{8}O_2+O_3\right\rangle.
\end{flalign}
\end{small}
These are renormalized multiplicatively without mixing. The renormalization constants correspond to the eigenvalues of $Z$.
\begin{small}
\begin{flalign}
&\left\langle O_{1\text{new}}\right\rangle =\left(1+\frac{3N\alpha _s}{4\pi \epsilon }\right)\left\langle O^{0}_{1\text{newB}}\right\rangle
\\ &\left\langle O_{2\text{new}}\right\rangle =\left(1+\frac{\left(15-\sqrt{17}\right)N\alpha _s}{48\pi \epsilon }\right)\left\langle O^{0}_{2\text{newB}}\right\rangle
\\ &\left\langle O_{3\text{new}}\right\rangle =\left(1+\frac{\left(15+\sqrt{17}\right)N\alpha _s}{48\pi \epsilon }\right)\left\langle O^{0}_{3\text{newB}}\right\rangle.
\end{flalign}
\end{small}
$O^{0}_{\text{newB}}$ means bare operator with bare fields and coupling. Finally, we can obtain the scale invariant condensates at the one loop order by multiplying these operators with corresponding factors of the coupling $\alpha_s$ so that the renormalization of the coupling cancels that of the operator \cite{Tarrach}.
\begin{flalign}
&\phi _1=\alpha _s{}^{-\frac{9}{11}}\left\langle O_{1\text{new}}\right\rangle
\\ &\phi _2=\alpha _s{}^{-\frac{15-\sqrt{17}}{44}}\left\langle O_{2\text{new}}\right\rangle
\\ &\phi _3=\alpha _s{}^{-\frac{15+\sqrt{17}}{44}}\left\langle O_{3\text{new}}\right\rangle.
\end{flalign}

\section{Summary}

We have identified and calculated the renormalization of the dimension 6 twist 4 gluon operators  to one loop order in the pure gauge theory.  Among the three independent operators,
$O_1$ is related to the second moment of the usual dimension 4 gluon condensate and do not mix with other operators $O_2$ and $O_3$, which vanishes in the pure gauge theory.  Hence, $O_1$ could be the first operator that can be estimated in a non perturbative model or calculated on the lattice.   With our calculation, the renormalization of all the dimension 6 operators are now known.   The QCD sum rule methods for the heavy quark system in medium can now be systematically studied up to dimension 6 level.

\section*{Acknowledgements}

This work was supported by the Korean Research Foundation under Grant Nos. KRF-2011-0020333 and KRF-2011-0030621.


\begin{thebibliography}{50}

\bibitem{Boyd:1996bx}
  G.~Boyd, J.~Engels, F.~Karsch, E.~Laermann, C.~Legeland, M.~Lutgemeier and B.~Petersson,
  Nucl.\ Phys.\ B {\bf 469}, 419 (1996)
  [hep-lat/9602007].

\bibitem{Furnstahl:1989ji}
  R.~J.~Furnstahl, T.~Hatsuda and S.~H.~Lee,
  Phys.\ Rev.\ D {\bf 42}, 1744 (1990).

\bibitem{Morita:2007pt}
  K.~Morita and S.~H.~Lee,
  Phys.\ Rev.\ Lett.\  {\bf 100}, 022301 (2008)
  [arXiv:0704.2021 [nucl-th]].

\bibitem{Gubler:2011ua}
  P.~Gubler, K.~Morita and M.~Oka,
  Phys.\ Rev.\ Lett.\  {\bf 107}, 092003 (2011)
  [arXiv:1104.4436 [hep-ph]].



\bibitem{Lee:1989qj}
  S.~H.~Lee,
  Phys.\ Rev.\ D {\bf 40}, 2484 (1989).

\bibitem{Lee:2008xp}
  S.~H.~Lee and K.~Morita,
  Phys.\ Rev.\ D {\bf 79}, 011501 (2009)
  [arXiv:0802.4000 [hep-ph]].

\bibitem{Manousakis:1986jh}
  E.~Manousakis and J.~Polonyi,
  Phys.\ Rev.\ Lett.\  {\bf 58}, 847 (1987).

\bibitem{Lee:2013dca}
  S.~H.~Lee, K.~Morita, T.~Song and C.~M.~Ko,
  Phys.\ Rev.\ D {\bf 89}, no. 9, 094015 (2014)
  [arXiv:1304.4092 [nucl-th]].

\bibitem{Klingl:1998sr}
  F.~Klingl, S.~s.~Kim, S.~H.~Lee, P.~Morath and W.~Weise,
  Phys.\ Rev.\ Lett.\  {\bf 82}, 3396 (1999)
  [Erratum-ibid.\  {\bf 83}, 4224 (1999)]
  [nucl-th/9811070].

\bibitem{Narison:1983kn}
  S.~Narison and R.~Tarrach,
  Phys.\ Lett.\ B {\bf 125}, 217 (1983).

\bibitem{Tarrach}
R.~Tarrach,
Nucl.\ Phys.\ {\bf B196}, 45 (1982)

\bibitem{SuHoungLee}
 S.~H.~Lee and S.~S.~Kim,
  Nucl.\ Phys.\  {\bf A679}, 517 (2001).

\bibitem{Abbott}
L.F.~Abbott,
Nucl.\ Phys.\ {\bf B185}, 189 (1981).



\end{thebibliography}
\end{document}